\PassOptionsToPackage{sort,compress}{natbib}
\documentclass[twocolumn,numberedappendix]{openjournal}

\usepackage{hyperref}
\hypersetup{
    colorlinks=true,
    linkcolor=blue,
    filecolor=blue,      
    urlcolor=gray,
    citecolor=blue,
}
\usepackage{xcolor}
\usepackage{textgreek}
\usepackage[utf8]{inputenc}
\usepackage[english]{babel}
\usepackage{color,colortbl}
\usepackage{tensind}
\tensordelimiter{?}
\DeclareGraphicsExtensions{.bmp,.png,.jpg,.pdf}
\usepackage{graphicx}

\usepackage{subcaption}

\usepackage{verbatim}
\usepackage[normalem]{ulem}
\usepackage{orcidlink}
\usepackage{soul}
\usepackage{bm}
\usepackage{amsmath}
\usepackage{float}
\usepackage{color}
\usepackage{xcolor}
\usepackage{booktabs}
\usepackage{cleveref}

\begin{document}

\title{High-Dimensional Bayesian Model Comparison in Cosmology with GPU-accelerated Nested Sampling and Neural Emulators}

\author{Toby Lovick$^{1}$, David Yallup$^{1,2}$, Davide Piras$^{3}$, Alessio Spurio Mancini$^{4}$, Will Handley$^{1,2}$}

\email{tcl44@cam.ac.uk}
\affiliation{1. Institute of Astronomy, University of Cambridge, Cambridge, UK}
\affiliation{2. Kavli Institute for Cosmology, University of Cambridge, Madingley Road, Cambridge, UK}
\affiliation{3. Département de Physique Théorique, Université de Genève, 24 quai Ernest Ansermet, 1211 Genève 4, Switzerland}
\affiliation{4. Department of Physics, Royal Holloway, University of London, Egham Hill, Egham, TW20 0EX, United Kingdom}

\begin{abstract}
We demonstrate a GPU-accelerated nested sampling framework for efficient high-dimensional Bayesian inference in cosmology. Using JAX-based neural emulators and likelihoods for cosmic microwave background and cosmic shear analyses, our approach provides parameter constraints and direct calculation of Bayesian evidence. In a 39-dimensional $\Lambda$CDM vs dynamical dark energy cosmic shear analysis, we produce Bayes factors and a robust error bar in just two days on a single A100 GPU, without loss of accuracy. Where CPU-based nested sampling can now be outpaced by methods relying on Markov Chain Monte Carlo (MCMC) sampling and decoupled evidence estimation, we demonstrate that with GPU acceleration nested sampling offers the necessary speed-up to put it on equal computational footing with these methods, especially where reliable model comparison is paramount. We also explore interpolation in the matter power spectrum for cosmic shear analysis, finding a further factor of 4 speed-up with consistent posterior contours and Bayes factor. We put forward both nested and gradient-based sampling as useful tools for the modern cosmologist, where cutting-edge inference pipelines and modern hardware can yield orders of magnitude improvements in computation time.
\end{abstract}

\section{Introduction}  
\label{sec:introduction}
Bayesian inference offers a principled statistical framework for parameter estimation and model comparison, and is widely employed in astrophysics for distinguishing between competing cosmological theories \citep{Trotta_2008}. The advent of large-scale surveys such as \textit{Euclid} \citep{Euclid_2025}\footnote{\href{https://www.euclid-ec.org/}{https://www.euclid-ec.org/}}, the Vera C. Rubin Observatory \citep{Vera}\footnote{\href{https://www.lsst.org/}{https://www.lsst.org/}}, and the Nancy Grace \textit{Roman} Space Telescope \citep{nancy}\footnote{\href{https://roman.gsfc.nasa.gov/}{https://roman.gsfc.nasa.gov/}}, alongside increasingly sophisticated theoretical models, has led to a significant increase in both data volume and model dimensionality. While parameter estimation remains manageable in this new scale of problem, the accurate computation of the Bayesian evidence stands as a distinct computational hurdle. Bayesian evidence is very valuable as a model comparison tool, as it weighs goodness of fit against model complexity, naturally penalising over-fit models \citep{Trotta07, Lovick:2023tnv}.

To address these high-dimensional inference challenges, the field has undergone significant algorithmic innovation. Advanced Markov chain Monte Carlo (MCMC) techniques, including gradient-based samplers like Hamiltonian Monte Carlo \citep[HMC;][]{Duane87, Neal96} and the No-U-Turn Sampler \citep[NUTS;][]{hoffman14a} which utilise differentiability for efficiency, Variational Inference (VI) offering an optimisation-based alternative \citep{hoffmanVI,Blei_2017,VIacceleration}, and machine learning-augmented approaches, such as emulators \citep[e.g.][]{Albers_2019, Manrique_Yus_2019, Mootoovaloo_2020, Arico_2022, Mootoovaloo_2022, SpurioMancini22CosmoPower, Piras_2023, El_Gammal_2023, guenther_2024, Bonici_2024_capse, Bonici_2024,guenther_2025} and the learned harmonic mean estimator for decoupled Bayesian evidence estimation \citep{mcewen2023machinelearningassistedbayesian, polanska2024learnedharmonicmeanestimation, polanska2025learnedharmonicmeanestimation, lin2025savagedickeydensityratioestimation}, have emerged as promising methods. Concurrently, the parallel processing capabilities of Graphics Processing Units (GPUs) have been identified as an important development, accelerating likelihood evaluations and sample generation across diverse methodologies \citep{Gu_2022, GPUacc,metha}, and extending the scope of feasible analyses.

Nested Sampling \citep[NS;][]{Skilling06} is a notable sampling framework that offers simultaneous parameter inference and Bayesian evidence calculation, a key advantage for rigorous model selection, while also effectively exploring complex, multimodal posterior landscapes. Although traditional CPU-based NS implementations have encountered scalability limitations in high-dimensional settings \citep{Feroz2009MultiNest}, recent work to adapt NS to efficiently utilise GPU-hardware \citep{NSSyallup} has pushed the capabilities of the algorithm much further than its CPU counterpart. Our work demonstrates that GPU-accelerated NS is significantly faster than its CPU counterparts on various cosmological likelihoods, while maintaining accurate posterior inference.

This paper is structured as follows. In Section \ref{sec:methodology} we describe at high-level the sampling algorithm and likelihoods demonstrated in this work, in Section \ref{sec:results} we display the model evidences and computation times of analyses, then in Section \ref{sec:discussion} we discuss and unpack the relative performance of the analyses, as well as the advantages and disadvantages between nested sampling and HMC with decoupled evidence calculation.

\section{Methodology}
\label{sec:methodology}
\subsection{Nested Sampling}
Nested Sampling (NS) is a Monte Carlo method designed to draw posterior samples and calculate the Bayesian evidence, $\mathcal{Z}$, defined as the integral of the likelihood $\mathcal{L}$ over the prior distribution $\pi$ of the parameters $\theta$:
\begin{equation}
    \mathcal{Z}=\int \mathcal{L}(\theta) \pi(\theta) d\theta \ . \label{eq:evidence}
\end{equation}

The evidence can then be used in Bayesian model comparison, where the posterior probabilities of models can be calculated using:

\begin{equation}
\frac{P(M_0|\text{d})}{P(M_1|\text{d})} = B_{01}\frac{\pi(M_0)}{\pi(M_1)}  \ , \quad B_{01} = \frac{\mathcal{Z}_0}{\mathcal{Z}_1} \ ,
\end{equation}

where the Bayes Factor $B_{01}$ and describes the relative strength of different models as they fit the observed data \citep{jeffreys1961}.

The algorithm iteratively replaces the point with the lowest likelihood among a set of ``live points'' with a new point drawn from the prior, constrained to have a likelihood higher than the point being replaced. This process allows the evidence to be accumulated as a sum of likelihood values weighted by successively smaller prior volumes. NS is well established as one of the primary methods for estimating the Bayesian evidence of models in cosmology, appearing in many contemporary high profile results~\citep[e.g.][]{planck,Abbott_2022, kidslegacy}. For a full review of NS aimed at physical scientists, see \citealt{Ashton_2022}.

The key challenge of implementing NS is to efficiently draw samples from the prior within these progressively shrinking, hard likelihood boundaries. Slice sampling~\citep{neal_slice_2003} is an MCMC technique well-suited for this task, as it can adaptively sample from such constrained distributions without requiring manual tuning of proposal scales. Usage of slice sampling within nested sampling was popularised in \cite{Polychord2,polychord}. Slice sampling works by sampling uniformly from an auxiliary variable defining a ``slice'' under the likelihood surface, and then sampling the parameters from the prior surface subject to a hard likelihood constraint. The work of \cite{NSSyallup} provided a generic implementation of the nested sampling algorithm, as well as a tailored implementation of a slice sampling based algorithm that was designed specifically to be amenable to the massive parallelism opportunities of modern GPUs. The latter implementation is styled as \emph{Nested Slice Sampling} (NSS), and we adopt this specific algorithm in this work.

\subsection{Parallel sampling and the future of inference}\label{samplerparall}

There are two main mechanisms by which a sampling code can be accelerated on GPU hardware. Firstly by leveraging the massively parallel SIMD (Same Instruction Multiple Data) paradigm to distribute a calculation over many threads, and secondly by utilising gradient information to improve the mixing of the algorithm itself. The frameworks used to encode calculations on GPUs typically afford fast evaluation of gradients via \emph{automatic differentiation}~\citep{jax2018github}, so this information is typically readily available for use. GPU native inference paradigms, such as momentum gradient descent as used to train neural networks~\citep{kingma2017adammethodstochasticoptimization}, are so successful as they utilise both of these mechanisms. Leveraging both these mechanisms in scientific probabilistic inference codes optimally is more challenging, and is the focus of this work.

HMC is a common choice in many scientific analyses, as a high level algorithm this can exploit the fast evaluation of gradients for rapid convergence. Popular self-tuning variants of HMC, such as NUTS, have been established to work well on many Bayesian inference problems across a variety of scales. However, the large variation in walk length incurred in these schemes makes parallelisation of multiple chains much less profitable, with contemporary work seeking to derive robust alternatives that are more amenable to GPU hardware~\citep{hoffman_adaptive-mcmc_2021,hoffman_tuning-free_2022}. 

Nested sampling, despite its prominence in many cosmological applications, does not make widespread use of either gradients or SIMD parallelism in its popular implementations~\citep{polychord, buchner2021ultranestrobustgeneral, Speagle_2020, albert2020jaxnshighperformancenestedsampling}. Realising the full potential of gradients within the nested sampling algorithm for general problems is an active area of research~\citep{betancourt_nested_2011,Cai_2022,lemos_improving_2023} we leave for future investigation, noting that establishing benchmark sampling problems in differentiable frameworks is a useful by-product of this work. Nested sampling can be parallelised by selecting $k$ live points with the lowest likelihoods rather than just the lowest, and evolving them with the likelihood constraint in parallel. This adaptation does not introduce new approximations into the core NS algorithm itself; rather, it significantly enhances computational speed by harnessing parallel hardware capabilities. If the generation of new points can be vectorised or batched, this represents a substantial potential speed-up for GPU execution, and evolving many points at once should scale as $\mathcal{O}(1)$ with the number of points, up to the parallelisation limit of the GPU. In practice, \cite{NSSyallup} demonstrated that perfect scaling is hard to achieve, however we demonstrate in this work that it is possible to gain orders of magnitude speed-up over classic NS pipelines.

The primary computational bottleneck in this parallel scheme is the speed of bulk likelihood evaluations. This motivates the use of \texttt{JAX}-based likelihoods and end-to-end cosmological pipelines \citep[e.g.][]{jaxcosmo, Ruiz_Zapatero_2024, Balkenhol_2024, Bonici_2025, Reymond_2025}. Further acceleration can be obtained by replacing traditional Boltzmann solvers such as \texttt{CAMB} \citep{Lewis11} or \texttt{CLASS} \citep{Diego_Blas_2011} with differentiable implementations \citep[e.g.][]{DISCO}, which are designed to run efficiently on modern hardware. Ultimately, the largest speed-ups are achieved through emulation with neural-network surrogates such as \texttt{CosmoPower-JAX} \citep{Piras_2023}, a \texttt{JAX}-based extension of the \texttt{CosmoPower} emulator framework \citep{SpurioMancini22CosmoPower}, which fully exploits the parallel processing capabilities of GPUs. 
The NSS implementation is provided in the \texttt{blackjax}~\citep{cabezas_blackjax_2024} framework. As a native \texttt{JAX}~\citep{deepmind_deepmind_2020} \emph{compilable} sampling framework, this is well positioned to exploit the parallelisation opportunities present at the likelihood level, reducing key factors such as communication overhead between devices that can plague hardware accelerated workflows.

In contrast, MCMC-based methods can access GPU vectorisation by running multiple independent chains in parallel. The generation of a new sample within a single chain is an inherently sequential process, so its speed is limited by the evaluation time of the likelihood function. While running many chains in parallel is an effective use of the GPU, this strategy requires that each chain independently reaches convergence, so the need for a ``burn-in" period for each chain to discard its initial non-stationary samples \citep{hoffman14a} means that the efficiency does not scale as directly as batching an NS run. 

As the target of our analysis is performing model comparison, on top of using HMC (NUTS) to draw samples, we employ the \texttt{harmonic} framework \citep{mcewen2023machinelearningassistedbayesian,polanska2025learnedharmonicmeanestimation}, which trains a normalising flow on posterior samples in order to compute the evidence via the learned harmonic mean estimator. We utilise the same flow architecture and training procedure as in \citet{Piras_2024}, training the flow on 30\% of chains and using the remaining 70\% for inference to compute the evidence. In this work \texttt{polychord} provides the comparison for a CPU-based nested sampler.

Whilst the NS analyses we develop in this work will prove to be scalable and competitive with other state-of-the-art model comparison techniques, without gradients it is important to recognise that full \emph{field-level} inference in cosmology (e.g.~\citealp{Jasche_2013, Leclercq_2015, Jasche_2019, Lavaux_2019, Ramanah_2019, Porqueres_2021, Porqueres_2021_b, bayer2023fieldlevelinferencemicrocanonicallangevin, Andrews_2023, Loureiro_2023, Sellentin_2023, Stadler_2023, Stopyra_2023, spuriomancini2024fieldlevelcosmologicalmodelselection, babic2025straighteningrulerfieldlevelinference, Lanzieri_2025, Mcalpine_2025}; see also \citealp{Leclercq_2025} for a recent review), is beyond the scope of this current NS implementation. It is however important to explore the limitations of current techniques, and developing competitive strategies in consistent frameworks (free from potential constraints of legacy codes) is vital to pushing the frontiers of inference problems in the physical sciences.
\begin{table}[t]
\centering
\caption{Prior distributions and fiducial values for the parameters of the cosmic shear model, split into cosmological, baryonic, and nuisance parameters. All priors used are uniform ($\mathcal{U}$) or Gaussian ($\mathcal{N}$). All values and ranges in this table follow \cite{Piras_2023}.}
\label{tab:shearpriors}
\begin{tabular}{c c c}
\toprule
\textbf{Parameter} & \textbf{Prior Range} & \textbf{Fiducial Value} \\
\midrule

$\omega_{\rm{b}} = \omega_{\rm{b}} h^2$ & $\mathcal{U}(0.01875, 0.02625)$ & 0.02242\\
$\omega_\text{cdm} = \Omega_\text{cdm} h^2$ &  $\mathcal{U}(0.05, 0.255)$ & 0.11933 \\
$h$ &  $\mathcal{U}(0.64, 0.982)$& 0.6766 \\
$n_{\rm{s}}$ & $\mathcal{U}(0.84, 1.1)$ & 0.9665 \\
$\ln 10^{10}A_{\rm{s}}$ & $\mathcal{U}(1.61, 3.91)$ & 3.047 \\
$w_0$ &  $\mathcal{U}(-1.5, -0.5)$ & -1\\
$w_a$ &  $\mathcal{U}(-0.5, 0.5)$ & 0\\
\midrule
$c_\text{min}$ & $\mathcal{U}(2, 4)$ & 2.6 \\
$\eta_0$ & $\mathcal{U}(0.5, 1)$  & 0.7\\

\midrule

$A_{\text{IA},i}$  & $\mathcal{U}(-6, 6)$ & $1-0.1i$\\
$D_{z_i}$ & $\mathcal{N}(0, 0.01^2)$ & 0\\
$m_i$   & $\mathcal{N}(0.01, 0.02)$  & 0.01 \\
\bottomrule
\end{tabular}
\end{table}
\subsection{CMB and Cosmic Shear Analyses}
\begin{figure*}[ht]
    \centering
    \begin{subfigure}[b]{0.48\textwidth}
        \centering
        \includegraphics{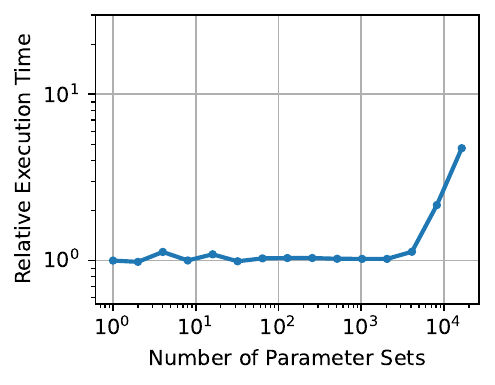}
        \caption{CMB likelihood as timed on an L4 GPU, with near-perfect vectorisation up to approximately $10^3$ parallel calls.}
        \label{fig:cmb_scaling}
    \end{subfigure}
    \hfill 
    \begin{subfigure}[b]{0.48\textwidth}
        \centering

        \includegraphics{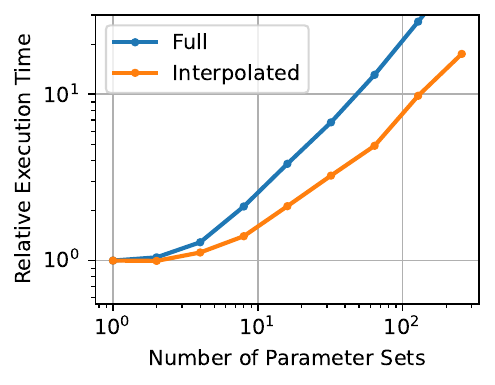} 
        \caption{Cosmic shear likelihood timed on an A100 GPU, where using interpolation introduces some vectorisation.}
        \label{fig:shear_scaling}
    \end{subfigure}
    \caption{Execution time for batched likelihood evaluations for both the CMB and cosmic shear likelihoods. Here the y-axis refers to the time relative to calling the likelihood on a single set of parameters. The interpolated likelihood involved evaluating the Matter Power Spectrum at fewer $z$-values, explained in detail in Appendix~\ref{appendix_interp}.}
    \label{fig:likelihood_scaling}
\end{figure*}

We demonstrate the sampler on two cosmological problems: a cosmic variance-limited CMB power spectrum analysis and a cosmic shear analysis. The CMB analysis, a standard 6-dimensional problem, functions as a benchmark and demonstration of the sampler's speed when used on a highly vectorisable likelihood, and its accuracy on a foundational application \citep{planck}. The cosmic shear analysis, with 7 cosmological and 30 nuisance parameters, presents a higher dimensional challenge that represents the edge of what CPU-based nested sampling can manage, and offers a direct comparison to the HMC + learned harmonic mean estimator analysis of \cite{Piras_2024}. Figure \ref{fig:likelihood_scaling} shows the vectorisation behaviour of both of these likelihoods, namely how they scale in execution time when called with batches of parameters at a time.

\subsection{CMB TT Power Spectrum}
The CMB temperature auto-power spectrum analysis is kept deliberately simple by assuming a cosmic variance-limited scenario. In this idealised case, the uncertainty in the power spectrum measurements $C_\ell^{\text{obs}}$ at each multipole $\ell$ is due only to the inherent statistical fluctuations of the CMB itself, rather than instrumental noise or foreground effects. The likelihood for the observed power spectrum, given a theoretical model $C_\ell(\theta)$ follows a chi-squared distribution of shape $2\ell +1$. The \texttt{CosmoPower-JAX} emulator suite provides the predictions for $C^{\rm{TT}}_\ell(\theta)$, allowing them to be vectorised  efficiently. This likelihood is run on mock data generated from the \textit{Planck} 2018 best-fit parameters \citep{planck}.

\subsection{Cosmic Shear}
We use the likelihood framework described in \citet{Piras_2023,Piras_2024}. Cosmic shear refers to the weak gravitational lensing effect where the observed shapes of distant galaxies are coherently distorted by the intervening large-scale structure of the Universe \citep{Kilbinger_2015}. This distortion is quantified by the angular shear power spectra $C_\ell^{\epsilon \epsilon}$.

The theoretical prediction for these $C_\ell^{\epsilon\epsilon}$ values involves integrating over the matter power spectrum $P_{\delta\delta}(k,z)$ and the lensing efficiency, which depends on the redshift distribution of source galaxies. \texttt{CosmoPower-JAX} is used to emulate the linear and non-linear matter power spectrum, and the $C_\ell^{\epsilon\epsilon}$ are calculated using the extended Limber approximation \citep{LoVerde_2008}, which translates between Fourier modes $k$ and multipoles $\ell$. Full details of the likelihood implementation and nuisance parameters are given in Appendix \ref{shearlikelihood}.

We consider two cosmological models for our model comparison, the fiducial $\Lambda$CDM cosmology, and the $w_0w_a$ parameterisation of evolving dark energy, also referred to as the CPL parameterisation \citep{CPL}. These models collectively have 5/7 cosmological parameters ($\Omega_{\rm{m}}, \Omega_{\rm{b}}, h, n_{\rm{s}}, A_{\rm{s}}$ and $w_0, w_a$) and 2 baryonic feedback parameters $c_\text{min}$ and $\eta_0$, which are used as the inputs for the emulator and cosmological functions, and then 30 nuisance parameters (3 for each of the 10 tomographic redshift bins), for a total of 37/39 dimensions. The prior ranges and fiducial values of each parameter are given in Table \ref{tab:shearpriors}.

To fully leverage the vectorisation capabilities of \texttt{JAX}, significant modifications were made to the original likelihood code. The original implementation using \texttt{JAX-cosmo} \citep{jaxcosmo}, built around `cosmology' class objects, offers high interpretability but is not amenable to \texttt{JAX}'s just-in-time (JIT) compilation of pure functions. We therefore refactored the likelihood into a series of stateless functions. This process resulted in a less general but highly optimised code path. Extensive validation was performed to ensure consistency between the two versions, with minor ($< 0.1\%$) numerical differences owing to slightly different choices of $\chi$ and $ D(\chi)$ implementations. By re-running HMC on our optimised likelihood we isolate the performance of each sampling method on the same likelihood function.

As shown in Fig.~\ref{fig:shear_scaling}, the full cosmic shear likelihood calculation is compute-intensive and does not scale efficiently beyond a small number of parallel evaluations. To explore the effects of alleviating this bottleneck, we also implement an alternative version of the likelihood, where the matter power spectrum emulator is evaluated over a smaller $z$-grid and interpolated. This trades accuracy for computational throughput and offers a comparison to the performance of a less computationally intensive likelihood. We report these results as the ``Full'' and ``Interpolated'' likelihoods.

\section{Results}
\label{sec:results}
We present the results of our comparative analyses in Table \ref{tab:results}, detailing the Bayesian evidence values and computation times for both the CMB and cosmic shear problems. For each problem, we compare our GPU-accelerated Nested Sampler (GPU-NS) against a traditional CPU-based NS implementation (\texttt{polychord}) and a modern HMC sampler coupled with a learned harmonic mean estimator (\texttt{harmonic}). We ran the GPU-NS with 1000 live points in both cases, and for HMC we ran 60 chains with 100/400 warm-up/samples and 400/2000 for the CMB and shear likelihoods respectively. This choice of 60 chains of 2000 samples, as well as all other tuning details of the HMC implementation, replicates \cite{Piras_2024}. In both cases the HMC and GPU-NS contours are in excellent agreement.

\subsection{CMB results}

\begin{table*}[ht]
\setlength{\tabcolsep}{4pt}
\caption{The evidence and computation times for the models and samplers considered, where ``wall-clock" times and ``GPU-hours" are quoted as the time to run \textit{both} cosmologies, and are given by their scale (days, hours, minutes) for interpretability. (I) denotes the interpolated likelihoods, and (*) denotes the times reported by \cite{Piras_2024}. The (*) models were both run on a \texttt{jax\textunderscore cosmo} implementation of the ``full" shear likelihood, where the \texttt{polychord} likelihood differs by a normalisation factor, as well as using \texttt{CAMB} instead of emulation for calculating the matter power spectrum. In this table, the GPU-hours and $\log B_{01}$ columns provide the clearest comparison between the NUTS+\texttt{harmonic} and GPU-NS methods.}
\label{tab:results} 
\vskip 0.15in
\begin{center}
\begin{small}

\begin{tabular}{llcccccc}
\toprule
Model  & Method & $\log \mathcal{Z}_{\Lambda\text{CDM}}$ & $\log \mathcal{Z}_{w_0w_a}$ & $\bm{\log B_{01}}$ & Wall Clock & Hardware & \textbf{GPU-Hours}  \\ 
\midrule
CMB  &  \texttt{polychord} & 87573.4$\pm$ 0.3& - &-&  1 hour  & 1 CPU & - \\
  &  GPU-NS & 87573.6$\pm$ 0.3& - & -& 12 seconds  & 1 L4 GPU & 0.003 \\
 &  NUTS + \texttt{harmonic} & 87572.3$\pm$ 0.007&- & -&  2 minutes  & 1 L4 GPU  & 0.03\\
\midrule
Shear &   GPU-NS &  $40958.09 \pm 0.31$ & $40955.49 \pm 0.33$ &$ 2.60 \pm 0.47$& 2 Days  & 1 A100 GPU & 48 \\
 &  NUTS + \texttt{harmonic} & $40956.58\pm  0.14$ &  $40954.38 \pm 0.31 $& $2.20 \pm 0.34 $ & 10 Hours & 1 A100 GPU & 10  \\
\midrule
Shear (I)  & GPU-NS &  $40959.03 \pm 0.31$ & $40956.28 \pm 0.32$ & $2.74 \pm 0.44$ & 11 hours  & 1 A100 GPU & 11 \\
 &  NUTS + \texttt{harmonic} & $40956.67\pm  0.38$ &  $40954.05 \pm 0.39 $& $2.62 \pm 0.55 $ & 6 Hours & 1 A100 GPU & 6  \\ \midrule
Shear (*)   &  \texttt{polychord} + \texttt{CAMB} & $-107.03\pm 0.27$& $-107.81 \pm 0.74$& $0.78 \pm 0.79$& 8 Months  & 48 CPUs & - \\
   &  NUTS + \texttt{harmonic} & $40956.55 \pm 0.06$& $40955.03 \pm 0.04$ & $1.53 \pm 0.07$ & 2 Days &12 A100 GPUs & 576\\ 

\bottomrule
\end{tabular}
\end{small}
\end{center}
\vskip -0.1in
\end{table*}
The CMB analysis highlights the ideal use-case for GPU-NS. As shown in Fig.~\ref{fig:cmb_scaling}, the emulator-based likelihood is highly vectorisable, and so our GPU-NS approach completes the analysis in just 12 seconds, a speed-up of nearly 300x compared to the 1 hour taken by \texttt{polychord}. While \texttt{polychord} updates a single live point per iteration, our method evolves 500 live points in parallel, fully exploiting the GPU's architecture. This performance gain would be even more pronounced against traditional, non-emulated likelihoods based on codes like \texttt{CAMB} \citep{Lewis:1999bs, Lewis11, Howlett:2012mh} or \texttt{CLASS} \citep{Lesgourgues_2011_a, Lesgourgues_2011_b,  Diego_Blas_2011}, which are significantly slower per evaluation.

The comparison with HMC in this regime is very instructive. While HMC is also fast, taking only $\sim2$ minutes, it does not match the raw speed of GPU-NS. This is because the primary advantage of HMC, namely its ability to take large, efficient steps using gradient information, is not as impactful as being able to vectorise the sampling en-masse. In this scenario, the parallelisation efficiency of the NS algorithm allows it to brute-force the calculation more rapidly than the multi-chain HMC approach.

\subsection{Cosmic Shear Results}
In the 37/39-dimensional cosmic shear analysis our GPU-NS framework transforms an analysis that is practically infeasible with CPU-based NS ($\sim$8 months reported in \citealp{Piras_2024}) into a task that can be completed by nested sampling in approximately 2 days on a single A100 GPU. The cosmological posteriors are shown in Fig.~\ref{shearmarg}, and the full 39-dimensional posterior is shown in Appendix~\ref{fullshearconstraints}. However, in contrast to the massive parallelism of the CMB-only analysis, Fig.~\ref{fig:shear_scaling} demonstrates that the more computationally intensive shear likelihood inhibits the potential speed-up achievable on a single GPU. Here HMC overtakes nested sampling, as it can use gradient information to explore the higher dimensional landscape just as well as the CMB likelihood. 

The interpolated likelihood achieves a factor of 4 speed-up over the full likelihood, and the Bayes factor and posterior contours (shown in Appendix~\ref{appendix_interp}) are in complete agreement with the full likelihood. Since the matter power spectrum is concave when plotted against $z$, interpolating causes a systematic sub-percent underestimate of the power spectrum, which is the source of the evidence differences between the interpolated and full likelihoods, however this appears to affect $\Lambda$CDM and $w_0w_a$ equally. Interpolating like this, while reliable for fiducial cosmological models, may miss small non-linear effects in the power spectrum. This speed-up therefore may not be accessible when probing more exotic physics.

In our posterior chains we note an additional, subdominant peak in the posterior around $A_{\text{IA},10} \sim 4.5$, which is unphysical and far from the fiducial value of $0$ (which the data was generated with). We stress this is a mock analysis and the actual likelihood in future analyses of real observations may take a different form, and that a full investigation of the impact of nuisance parameters is beyond the scope of this paper.

\section{Discussion}
\label{sec:discussion}

\begin{figure*}
    \centering
    \includegraphics[width=0.7\linewidth]{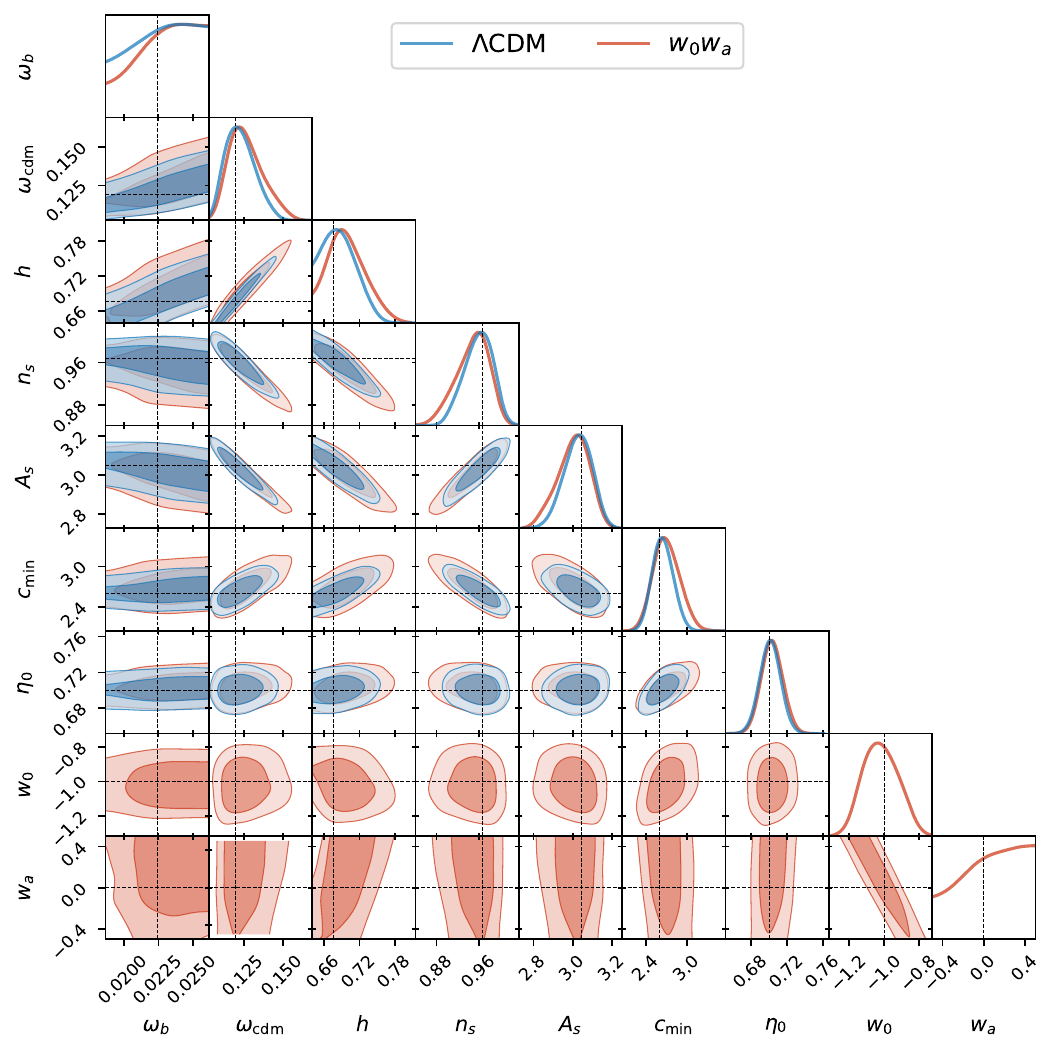}
    \caption{Marginalised posterior distributions for the cosmological parameters of the cosmic shear analysis, using the full likelihoods with GPU-NS. The contours show the 68\% and 95\% credible intervals for each parameter. The dashed lines show the fiducial values that the mock data was generated with.}
    \label{shearmarg}
\end{figure*}

\subsection{Performance breakdown}
On the CMB analysis, where as shown in Fig.~\ref{fig:cmb_scaling} the likelihood perfectly vectorises well beyond our chosen resolution of 1000 live points, nested sampling finds its biggest speed-up. As presented in \cite{NSSyallup}, the \texttt{blackjax} nested sampling implementation is a well-vectorised sampling procedure, and thus can fully leverage parallel evaluation of this likelihood. 

While it is feasible to gain this level of speed-up with HMC on a vectorised likelihood, its statistical validity relies on using a few long chains rather than many short ones. MCMC requires chains to be sufficiently long to both discard an initial warm-up sequence and to generate samples with low auto-correlation. A sampling strategy of 500 chains of 20 samples would be statistically poor; it would be dominated by warm-up overhead and the resulting chains would be too short to effectively explore the parameter space. This is in contrast to a run of 10 chains with 1000 samples each, which provides a much more robust estimate of the posterior, but is only able to calculate 10 likelihood evaluations in parallel.

When the likelihood is the bottleneck, such as in the cosmic shear analysis, GPU-NS loses its advantage, and the two algorithms sit on a fairly equal footing. Here all of the speed-up comes from the likelihood internally vectorising:although the samplers cannot massively vectorise to accelerate the inference, individual likelihood calls are greatly sped up by having their many internal emulator calls vectorised. The interpolated likelihood results are very instructive, because the change affects the speed of the GPU-NS and HMC analyses differently. While HMC runs twice as fast on the interpolated likelihood, GPU-NS sees a speed-up of more than four times. This is likely from it better utilising the increased vectorisation shown in Figure~\ref{fig:shear_scaling}, which comes from reducing the number of emulator calls per likelihood. This, along with the CMB timing results, supports that GPU-NS is very efficient when run on vectorised likelihoods.

There are a number of potential routes to lift this bottleneck further. All the calculations presented in this work are performed in double numeric precision: whilst it is non-trivial to ensure that sufficient accuracy is achievable with lower precision, it is noteworthy that much of the hardware development is targeting increasingly reduced numeric precision, and this is a development that physicists need to be alert to~\citep{jacob2017quantizationtrainingneuralnetworks}. Reducing the required numeric precision to single float numerics would greatly increase the potential to parallelise the calculations in this work. More immediately, it is also feasible to recover massive speed-ups by deploying the calculations over multiple devices/GPU clusters. The original work of ~\citet{Piras_2024} distributed its MCMC chains over multiple devices, thus enabling faster (in wall-time) accumulation of multiple chains. However, multiple chains in this context primarily improve the estimate of the variance of the result; in contrast, population level approaches such as NSS can use multiple devices to directly improve run-time (at a fixed variance). We leave a multi-device implementation of this pipeline to future work.

\subsection{Comparison of Methods}
This work shows that with GPU-acceleration nested sampling is now feasible on high-dimensional analyses such as the cosmic shear analysis presented here. However, it should be noted that although we have provided enough speed-up to explore a 39-dimensional prior, the number of likelihood evaluations nested sampling needs to converge scales very harshly with dimension \citep{polychord}, and as such field-level nested sampling and other ultra-high-dimensional problems are still out of reach, even with these accelerations. In comparison HMC, due to its usage of gradient information, is able to explore posteriors in much higher dimensions and has already been used in field-level inference. This is a significant drawback of nested sampling, with \texttt{JAX}-based pipelines all of our models are auto-differentiable, information that at the present is not being used by nested sampling.

Some caution should be taken with the error bars provided by the learned harmonic mean estimator, as it represents only the scatter in the estimate taken on different subsets of the data. It does not include any error from how the flow might change if trained on a different set of posterior samples, and most importantly there is no systematic error from the flow not capturing the true posterior shape. This incomplete error budget is a concern: the GPU-NS versus HMC results presented in Table \ref{tab:results} are around $4\sigma$ away from each other for both the full and interpolated likelihoods; so while each method appears to recover the same Bayes factor, a discrepancy of 2 in log space is the difference between decisive and inconclusive evidence when performing Bayesian model comparison \citep{jeffreys1961}, so models certainly cannot be compared across methods (GPU-NS against HMC), even on identical likelihoods. It should be noted that the discrepancies between the Bayes factors of this work and of \cite{Piras_2024} may have arisen due to the minor numerical differences in the likelihood implementations mentioned before, so should not be compared as directly as our HMC and GPU-NS results.

Nested Sampling, in turn, is not immune to its own implementation-specific biases, and its error bar is also conditional. The NS error calculation assumes that at each iteration, a new live point is drawn perfectly uniformly from the prior volume constrained by the current likelihood threshold \citep{Skilling06}. In practice, this is approximated with an inner MCMC sampler which, in high-dimensional or complex posteriors, can struggle to explore the entire valid region. This inefficiency can introduce a systematic bias that the reported statistical error does not account for. However, since nested sampling is a global algorithm (sampling over the entire prior instead of rapidly settling into the posterior), it can properly account for the error in the specific path taken when crossing from the prior to the posterior, and it is unlikely to miss modes when run at sufficient precision. The secondary $A_{\text{IA},10}$ peak, for example, was found in very few of the HMC chains, due to the sampler's sensitivity to its initialisation, and was thus under-represented by those posterior samples. We stress that this is an unphysical mode in a nuisance parameter, and may well be just an artefact of the formulation of the likelihood.

Bringing the overall runtime of this challenging frontier problem to a level that multiple methods can be tested with rapid turn around is a significant success of the optimisations performed as part of this work. It is now possible to investigate the discrepancy between the \texttt{harmonic} and NS based pipelines by incorporating other advanced MCMC-based normalising constant estimators such as Sequential Monte Carlo~\citep{doucet_introduction_2001, chopin2020introduction}. 

We claim that maintaining and developing multiple approaches to likelihood-based inference is essential for consistency and robustness. Both methods have a rightful place in the modern cosmologist’s toolkit, and our discussion illustrates their respective strengths and weaknesses, particularly for problems at this scale.

\section{Conclusion}

We have demonstrated that the combination of GPU-acceleration, \texttt{JAX}-based emulators, and a vectorised Nested Sampling algorithm removes the primary computational barrier in using Nested Sampling in high-dimensional problems. Our framework achieves a speed-up of over two orders of magnitude on a cosmic-variance-only CMB analysis and, more critically, reduces the runtime of a 39-dimensional cosmic shear analysis from months to a matter of days, placing Nested Sampling on an equal computational footing with the fastest alternative methods.

The reason to speed-up current analyses is not just to re-run those problems faster; with current analyses now more feasible, we can be confident in our ability to analyse larger upcoming data sets in more detail, but also to perform wider searches over model space. An analysis that is orders of magnitude faster allows us to test a whole grid of foreground and beyond-$\Lambda$CDM cosmological models on a given experiment with the same computational resources as before.

Nested sampling is uniquely positioned within the field of cosmology as a reliable baseline for accurate evidence calculation; however, its implementation is typically tied to legacy code pipelines. This work establishes that, through the use of algorithms and forward models built to leverage modern hardware, nested sampling can be a routine component of the next generation of cosmological analyses. As we move towards future experiments, we must be mindful of trading off precision in order to make analyses computationally feasible, and this work demonstrates that nested sampling will continue to be a viable tool in the next stage of cosmological surveys and experiments.

While this and other work \citep{NSSyallup,metha} establishes GPU nested sampling as being able to greatly accelerate medium- to high-dimensional analysis, future work will push the scale of feasible analyses even further, now made possible by building our inference pipelines around GPU hardware. Developments such as efficient parallelisation of multi-GPU instances, as well as efforts at pushing emulation even further, will enable the acceleration of the next tier of problems.

\section*{Acknowledgements}
This work was supported by the research environment and infrastructure of the Handley Lab at the University of Cambridge. TL is supported by the Harding Distinguished Postgraduate Scholars Programme (HDPSP). DP acknowledges financial support from the Swiss National Science Foundation.

\section*{Data Availability}

The nested sampling chains and HMC posterior samples underlying this article are available on Zenodo, at https://doi.org/10.5281/zenodo.17131319

\bibliography{GPUNS}

\appendix

\section{Cosmic Shear Likelihood Details}\label{shearlikelihood}
We consider a tomographic survey with $N_{\text{bins}}=10$ redshift bins. The primary observable is the angular power spectrum of the cosmic shear signal $C^{\epsilon \epsilon}_{i j} (\ell)$ between all pairs of bins $(i,j)$. This signal is composed of the true cosmic shear signal ($\gamma$) and a contaminant from the intrinsic alignments ($I$) of galaxies:
\begin{equation}
	C^{\epsilon \epsilon}_{i j} (\ell) = C^{\gamma \gamma}_{i j} (\ell) + C^{\gamma I}_{i j} (\ell) + C^{I \gamma}_{i j} (\ell) + C^{I I}_{i j} (\ell) \ .
\end{equation}
Each of these components is computed using the extended Limber approximation \citep{LoVerde_2008}, which projects the 3D matter power spectrum, $P_{\delta \delta}(k,z)$, into a 2D angular correlation:
\begin{equation}
	C^{A B}_{i j} (\ell) = \int_0^{\chi_H} \frac{W_i^A (\chi)  W_j^B (\chi)}{\chi^2}  P_{\delta \delta} \left(k = \frac{\ell + 1/2}{\chi}, z \right) \mathrm{d}\chi \ ,
\end{equation}
where $\{A, B\} \in \{\gamma, I \}$, $\chi$ is the comoving distance integrated up to the horizon $\chi_H$, and $W(\chi)$ are the window functions for each component.

The shear window function $W^\gamma(\chi)$ depends on the underlying cosmology and the redshift distribution $n_i(z)$ of source galaxies in each tomographic bin $i$:
\begin{equation}
	W_i^\gamma (\chi) = \frac{3 \Omega_{\rm m} H_0^2 }{2  c^2} \frac{\chi}{a(\chi)} \int_{\chi}^{\chi_{\rm H}}   n_{i}(\chi')  \frac{\chi'-\chi}{\chi'} \mathrm{d} \chi' \ ,
\end{equation}
which is sensitive to the matter density $\omega_{\rm{m}}$, the Hubble parameter $H_0$ and the scale factor $a$. To achieve the precision required by modern surveys, we must also model several systematic effects, each of which introduces nuisance parameters that must be marginalised over.
\begin{itemize}
    \item \textbf{Intrinsic Alignments ($A_{\text{IA},i}$):} The intrinsic shapes of nearby galaxies can be physically correlated, mimicking a lensing signal. We model this using the Non-Linear Alignment (NLA) model \citep{hirata_04} as parametrised in \citet{Piras_2023}, which defines the intrinsic alignment window function $W^I$. We introduce a single free amplitude parameter per bin, $A_{\text{IA},i}$, controlling the strength of the signal:
    \begin{equation}
    	W_i^{\rm I}(\chi) = - A_{\mathrm{IA}, i}  \frac{C_1  \rho_{\rm cr}  \Omega_{\rm m}}{D(\chi)}  n_{i}(\chi) \ ,
    \end{equation}  
     which is sensitive to the linear growth factor $D(\chi)$, the critical density of the universe $\rho_{\rm cr}$, and a normalisation constant $C_1$ fixed to $5 \times 10^{-14} \, h^{-2} M_\odot^{-1} \text{Mpc}^3$.     This adds 10 nuisance parameters to our model.
    \item \textbf{Multiplicative Shear Bias ($m_i$):} Imperfections in the shape measurement algorithms can lead to a systematic bias in the measured shear, which we model with a multiplicative factor $m_i$ for each bin. The observed shear spectrum is rescaled by $(1 + m_i)(1 + m_j)$. This introduces another 10 nuisance parameters.

    \item \textbf{Photometric Redshift Errors ($D_{z_i}$):} Uncertainties in photometric redshift estimation can shift galaxies between bins. We model this with a simple shift parameter $D_{z_i}$ for the mean of each redshift distribution, $n_i(z) \rightarrow n_i(z - D_{z_i})$ \citep{Eifler}, adding a final 10 nuisance parameters.
\end{itemize}

    The likelihood for the data vector of all angular power spectra is assumed to be a multivariate Gaussian, with a simulated covariance matrix for an LSST-like survey ($n_{\text{gal}} = 30\,\text{arcmin}^{-2}$, $\sigma_{\epsilon} = 0.3$, $f_\text{sky} = 0.35$). The $C_{ij}^{AB}$ and $W$ integrals are all evaluated on a grid of $1025$  $\chi$ values that are evenly spaced in scale factor from $a=1$ down to the furthest object in the mock data, $a_\text{min} = (1+5.02)^{-1} = 0.16$.

The three likelihoods mentioned in Table \ref{tab:results}, namely Shear, Shear (I), and Shear (*), are differentiated by their exact implementation of the $C^{\epsilon \epsilon}_{i j} (\ell)$. (*) is the original \texttt{jax\textunderscore cosmo} implementation of the likelihood as used in \cite{Piras_2024}. The main shear likelihood of this paper is the refactored version of (*) that aimed to be a more \texttt{JAX}-native and optimised version of that code, while Shear (I) is the interpolated version of this optimised likelihood.
\newpage
\section{Accuracy of Interpolated Likelihoods}\label{appendix_interp}
\begin{figure}[h]
    \centering
    \includegraphics[width=0.75\textwidth]{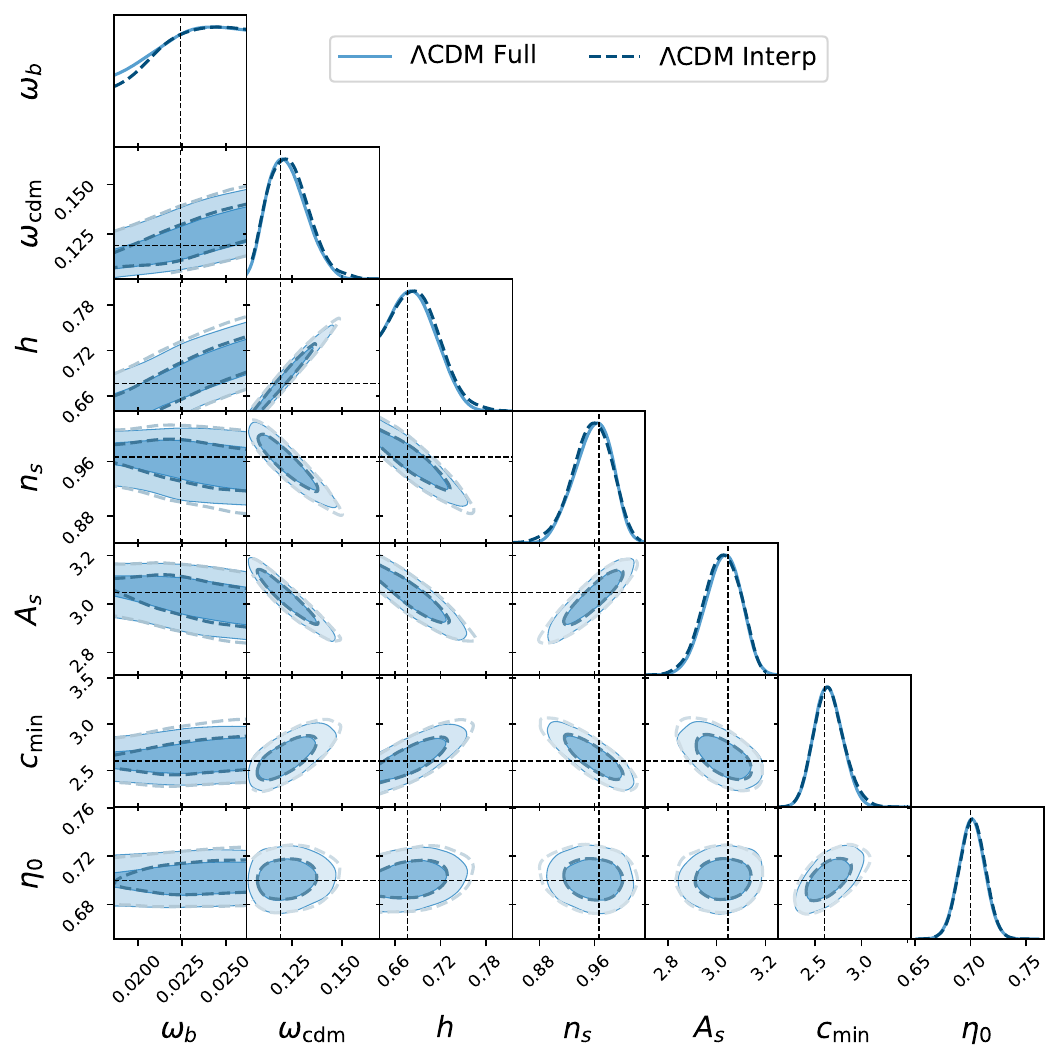}
    \caption{A comparison of the NS posterior contours of the $\Lambda$CDM likelihood with and without interpolation over the matter power spectrum. The posteriors are in excellent agreement, although the likelihood values of each model differ by a systematic $\Delta \log \mathcal{L} \approx 2$.}
    \label{interpversusfull}
\end{figure}

The interpolated likelihood (I) evaluates the matter power spectrum on a redshift grid of 10 times lower resolution than the other likelihoods, and then linearly interpolates over $z$ before evaluating the $k$-modes. For this likelihood all integrals are still performed on the full grid of 1025 redshift/scale factor/$\chi$ values. Figure \ref{interpversusfull} shows excellent agreement between the full and interpolated posteriors, despite the relative evidence differences between the methods as shown in Table \ref{tab:results}. As mentioned in Section \ref{sec:results}, the evidence differences between the two likelihoods appears to be constant over parameter space, resulting from a systematic underestimate of $P_{\delta\delta}$ values when interpolating the concave matter power spectrum.
\newpage
\section{Full Cosmic Shear Constraints}\label{fullshearconstraints}

\begin{figure}[H]
    \centering
    \includegraphics[width=\textwidth]{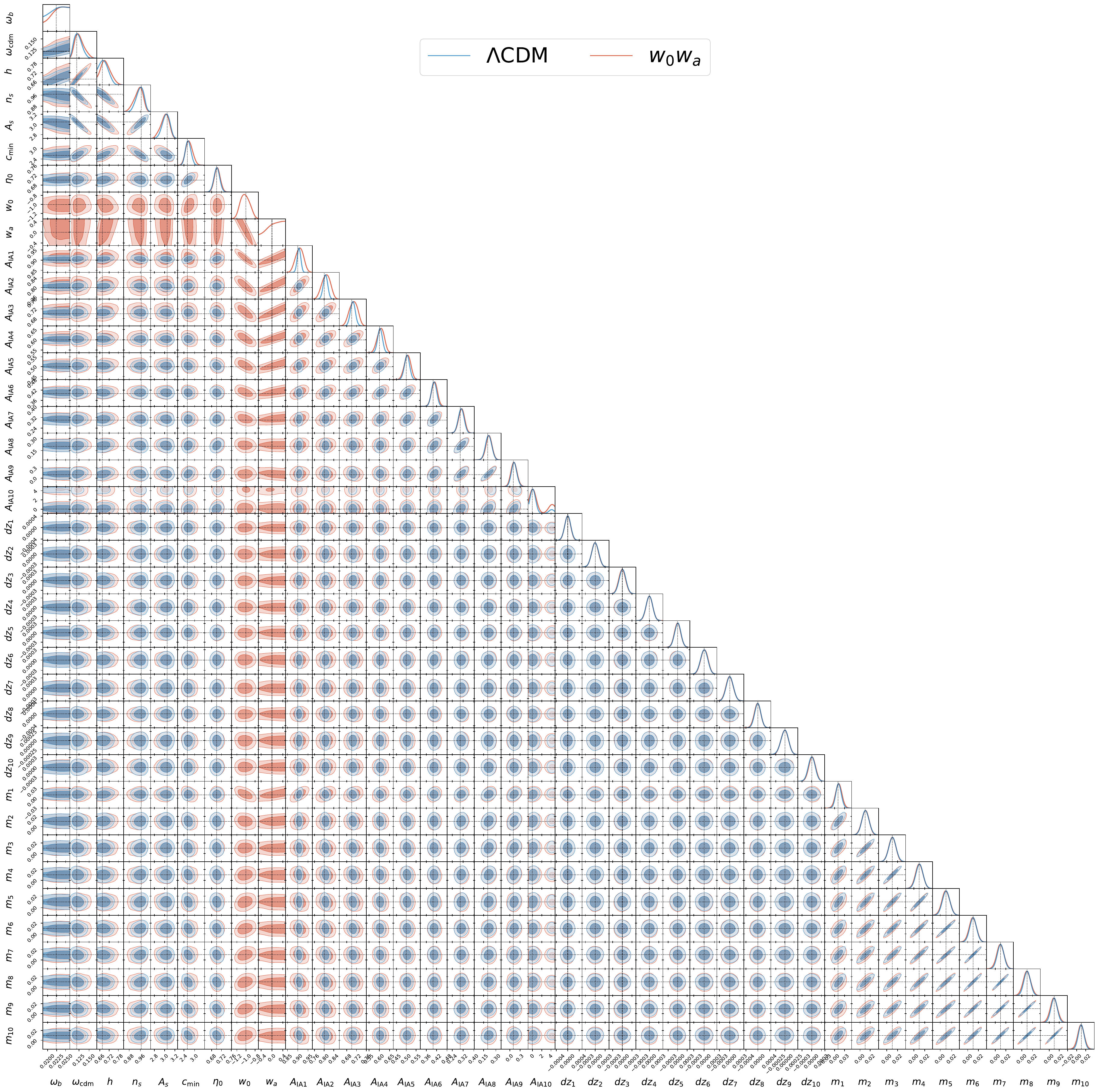}
    \caption{Full posterior constraints for the 39 parameters of the $\Lambda$CDM (blue) and $w_0w_a$(orange) cosmologies from cosmic shear , obtained with GPU-NS. The dashed lines show the truth values of the parameters which were used to generate the data.}
    \label{shearfull}
\end{figure}

For completeness, we present the full posterior constraints from the cosmic shear analysis in Figure \ref{shearfull}. This figure displays the marginalised one- and two-dimensional posterior distributions for all parameters, comparing the 39-dimensional $w_0w_a$ model against the nested 37-dimensional $\Lambda$CDM model.

\end{document}